\documentclass[12pt]{article}

\usepackage{amsmath}
\usepackage{amsfonts}
\usepackage{amssymb}
\usepackage{graphicx}
\usepackage{psfrag}
\usepackage{graphicx}
\usepackage{psfrag}

\newcommand{\be}{\begin{equation}}
\newcommand{\ee}{\end{equation}}
\newcommand{\ba}{\begin{eqnarray}}
\newcommand{\ea}{\end{eqnarray}}

\begin{document}
\begin{center}
{\bf New solutions for graphene with scalar potentials by means of generalized intertwining}\\
\vspace{1cm}
M. V. Iof\/fe$^{1,}$\footnote{E-mail: ioffe000@gmail.com; corresponding author},
D. N. Nishnianidze$^{2,}$\footnote{E-mail: cutaisi@yahoo.com}, E. V. Prokhvatilov$^{1,}$\footnote{E-mail: evgeniprokhvat@yandex.ru}\\
\vspace{0.5cm}
$^1$ Saint Petersburg State University, 7/9 Universitetskaya nab., St.Petersburg, 199034 Russia.\\
$^2$ Akaki Tsereteli State University, 4600 Kutaisi, Georgia.\\

\end{center}
\vspace{0.5cm}
\hspace*{0.5in}
\vspace{1cm}
\hspace*{0.5in}
\begin{minipage}{5.0in}
{\small
The intertwining relations between superpartner Hamiltonians are the main ingredients of well known Supersymmetrical Quantum Mechanics (SUSY QM). In the present paper, the generalized form of intertwining is used for the investigation of a massless (zero energy) two-dimensional Dirac equation with scalar external potential. This equation is related to the description of graphene and some other materials in the field of external electrostatic potential. The use of modified intertwining relations allows to find analytically solutions for the wave functions in the field of some external scalar potentials which depend on both space coordinates. A few examples of this construction are given explicitly.
}
\end{minipage}

Keywords: intertwining relations, two-dimensional Dirac equation, graphene, supersymmetrical Quantum Mechanics.\\
{\it PACS:} 03.65.Fd; 73.22.Pr; 11.30.Pb; 03.65.-w;

\section{Introduction.}

During last decades, starting from the pioneer paper of Witten in 1981 \cite{witten}, the supersymmetrical approach became a new interesting and effective study method in investigation of various classical and modern problems of Quantum Mechanics. This approach combined the characteristic features of two old well known methods: Darboux Transformations in Mathematical Physics \cite{darboux}, \cite{darboux-2} and Factorization Method of Schr\"odinger \cite{schr}, \cite{schr-2} in Quantum Mechanics (see details in \cite{reviews} - \cite{reviews-5}). Their combination was expressed in the algebraic form of SUSY QM where both commutation and anticommutation relations connect three superoperators - Superhamiltonian $\hat H,$ and Supercharges $Q^+,\,Q^-:$
\be
\{Q^+, Q^-\}_+=\hat H;\quad [\hat H, Q^{\pm}]=0.
\label{super}
\ee
The superalgebra (\ref{super}) admits a lot of different realizations, and each of them corresponds to a specific kind of quantum systems such as models with different space dimensionality, stationary/non-stationary models, models with scalar/matrix interaction, systems with Hamiltonians of second/first order in momenta, discrete/continuous spectra, one/many-particle systems etc. \cite{nature} - \cite{nature-10}. The simplest possible realization of the algebra (\ref{super}) has the form:
\be
\hat H = \left(
                \begin{array}{cc}
                  h^{(0)} & 0 \\
                  0 & h^{(1)} \\
                \end{array}
              \right);\quad
Q^+=\left(
            \begin{array}{cc}
              0 & 0 \\
              q^- & 0 \\
            \end{array}
          \right);\quad
Q^-=(Q^+)^{\dagger}=\left(
            \begin{array}{cc}
              0 & q^+ \\
              0 & 0 \\
            \end{array}
          \right). \label{2times2}
\ee
In particular, such realization was used in the framework of standard one-dimensional stationary scalar Quantum Mechanics, where the partner Hermitian Hamiltonians $h^{(0)},\,h^{(1)}$ are the second order differential operators and the components $q^{\pm}$ of supercharge are of first order in momenta. The latter ones play a dual role: they simultaneously factorize Hamiltonians and intertwine them,
\ba
h^{(0)}&=&q^+q^-;\quad h^{(1)}=q^-q^+;\quad q^-=(q^+)^{\dagger};  \label{realization-1}\\
h^{(0)}q^+&=&q^+h^{(1)};\quad h^{(1)}q^-=q^-h^{(0)}. \label{realization-2}
\ea
Just the intertwining relations (\ref{realization-2}) provide the relations between wave functions and the corresponding spectra of two spectral problems - with Hamiltonians $h^{(0)}$ and $h^{(1)}$ \cite{reviews} - \cite{reviews-5}.

It is important that the intertwining relations (\ref{realization-2}) be considered separately when no factorization similar to (\ref{realization-1}) is fulfilled. For example, when both Hamiltonians and intertwining operators are differential operators of the same degree in derivatives \cite{acdi}, \cite{acdi-2}. Exactly in this sense, the intertwining relations play a significantly more important role in SUSY QM.

In the present paper, we will go further by modifying essentially the SUSY intertwining themselves. The intertwining operators $q^+,\,q^-$ in (\ref{realization-2}) will not be now the same in l.h.s and r.h.s., namely:
\be
 h^{(0)}q^+=\widetilde q^+h^{(1)};\quad h^{(1)}\widetilde q^-=q^-h^{(0)}, \label{realization-3}
\ee
with different $q's$ and $\widetilde q's$. From mathematical point of view, relations of similar form were studied in \cite{shemyakova}, \cite{shemyakova-2} but in different context. Specifically, we will study solutions of relations (\ref{realization-3}) with a pair of first order $2\times 2$ two-dimensional matrix differential operators $D_1, \, D_2$ of Dirac form (instead of Schr\"odinger Hamiltonians $h$ above) and also first order matrix intertwining operators $q^{\pm},\, \widetilde q^{\pm}.$

The motivation to use such a generalization is the following. While for the solution of the standard form of spectral problem $H\Psi_n=E_n\Psi_n$ with unknown eigenvalues $E_n$
and wave functions $\Psi_n,$ the usual intertwining relations (\ref{realization-2}) are adequate, in the case of searching for solutions of $D\Psi=0,$ the use of generalized intertwining of the form (\ref{realization-3}) is enough. Do such homogeneous equations describe any known physical system? Yes, just such equations - two-dimensional Dirac equations with zero mass in the field of external (scalar or matrix) potential - have to be studied in the framework of theory of graphene and some other materials \cite{novoselov} - \cite{novoselov-4}. The two-component solutions of these Dirac equations provide the probability distribution of electron carriers. Here, the up and down elements of the Dirac two-component spinor are the wave functions corresponding to two different sublattices in graphene. Such form of Dirac equation was studied in many papers, mainly with external potentials depending only on one spatial variable \cite{peres} - \cite{peres-12}.

Among others, the methods of (standard) SUSY Quantum Mechanics were also effectively used \cite{susy}, \cite{susy-2} for the case of one-dimensional potential but
recently in \cite{IN} for a class of external potentials, solutions with actual dependence on both coordinates were built analytically.
In the present paper, the problem with external potentials $V(\vec x)$ of a scalar (electrostatic) nature will be studied by means of generalized intertwining relations which should significantly increase the number of pairs of partner potentials. Among these potentials one can look for so simple that they are amenable to analytic solution of the corresponding Dirac equation. Then the partner Dirac equation can be also solved analytically due to generalized intertwining relations. The structure of the paper is the following. The next Section contains a detailed description of the generalized variant of SUSY intertwining relations for two-dimensional Dirac operators by means of matrix intertwining of first order in momenta. Several examples are collected in Section 3 where the solutions of both partner Dirac equations can be found analytically. In the conclusions, the results are summarized, and the role of zero modes of intertwining operator is discussed.

\section{Method of generalized intertwining relations.}

Let us consider the problem mentioned above of the analytical solution of two-dimensional massless (or, equivalently, zero energy) Dirac equation with a scalar potential:
\be
D_1\Psi^{(1)}(\vec x)=0;\quad D_1\equiv (-i\sigma_1\partial_1 - i\sigma_2\partial_2 + V_1(\vec x)), \quad \partial_i\equiv \frac{\partial}{\partial x_i},\, i=1,2,  \label{dirac}
\ee
where the Fermi velocity was taken to unity, $\sigma_1, \sigma_2$ - standard Pauli matrices, $x_1, x_2$ - spatial coordinates, $V_1(\vec x)$ - real scalar potential, and $\Psi^{(1)}(\vec x)$ is a two-component "spinor" with components $\Psi_A^{(1)}(\vec x), \Psi_B^{(1)}(\vec x).$ The main idea is to look for the partner Dirac operator $D_2$ which is intertwined with $D_1$ as
\be
D_1 N = M D_2;\quad D_2=-i\sigma_1\partial_1 - i\sigma_2\partial_2 + V_2(\vec x)
\label{int}
\ee
with two {\bf different} first order differential matrix operators:
\be
N(\vec x) \equiv A_n\partial_n + A(\vec x);\quad M(\vec x) \equiv B_k\partial_k + B(\vec x),
\label{NM}
\ee
with summation over repeated indices $n, k = 1,2.$ In (\ref{NM}), $A_n,\, B_k$ are constant $2\times 2$ matrices but matrices $A(\vec x),\, B(\vec x)$ may depend on coordinates. Let us suppose that the partner real potential $V_2(\vec x)$ is simple enough so that the two-component solutions $\Psi^{(2)}(\vec x)$ of equation $D_2\Psi^{(2)}(\vec x)=0$ can be found. In such a case, due to generalized intertwining (\ref{int}), solutions of Eq.(\ref{dirac}) can be built as well:
\be
\Psi^{(1)}(\vec x)= N(\vec x)\Psi^{(2)}(\vec x).
\label{solutions}
\ee
Thus, the problem can be reformulated as follows. One has to solve the generalized intertwining relations (\ref{int}), i.e. to find both potentials
$V_1(\vec x), \, V_2(\vec x),$ all constant matrices $A_n, B_k$ and matrix functions $A(\vec x), B(\vec x).$ To find solutions, one has to decompose
(\ref{int}) step by step over Pauli matrices and over different partial derivatives.

Equating the coefficients with second derivatives in (\ref{int}), we obtain:
\ba
&&A_1=a_0^{(1)}+\vec{a}^{(1)}\cdot \vec{\sigma} ;\nonumber\\
&&A_2=a_0^{(2)}-a_2^{(1)}\cdot\sigma_1+a_1^{(1)}\cdot\sigma_2+a_3^{(2)}\cdot\sigma_3; \label{a}\\
&&B_1=a_0^{(1)}+a_1^{(1)}\cdot\sigma_1-a_2^{(1)}\cdot\sigma_2-a_3^{(1)}\cdot\sigma_3; \nonumber\\ &&B_2=a_0^{(2)}+a_2^{(1)}\cdot\sigma_1+a_1^{(1)}\cdot\sigma_2-a_3^{(2)}\cdot\sigma_3,
\label{b}
\ea
where all $a's$ are constant coefficients.

Analogously, equating the coefficients in two relations of intertwining (\ref{int}) with first derivatives and using (\ref{a}) and (\ref{b}), one derives the relations between coefficients of expansions of (\ref{NM}),
\be
A(\vec x) \equiv a_0(\vec x)+\vec{a}(\vec x)\cdot\vec\sigma;\quad B(\vec x) \equiv b_0(\vec x)+\vec{b}(\vec x)\cdot\vec\sigma
\label{abab}
\ee
and combinations
\be
V_{\pm}(\vec x) \equiv V_1(\vec x) \pm V_2(\vec x).
\ee
Namely, these relations are:
\ba
2b_1(\vec x) &=& a_3^{(2)}V_+(\vec x) + i a_0^{(1)}V_-(\vec x); \quad 2b_2(\vec x) = ia_0^{(2)}V_-(\vec x) - a_3^{(1)}V_+(\vec x);
\nonumber\\
A(\vec x) &=& b_0(\vec x)-ia_1^{(1)}V_-(\vec x)+\frac{1}{2}(a_3^{(2)}V_+(\vec x)-ia_0^{(1)}V_-(\vec x))\cdot\sigma_1- \nonumber\\
&-&\frac{1}{2}(a_3^{(1)}V_+(\vec x)+ia_0^{(2)}V_-(\vec x))\cdot\sigma_2 + (a_2^{(1)}V_+(\vec x)-b_3(\vec x))\cdot\sigma_3.
\nonumber
\ea

The last step is to derive from (\ref{NM}) the relation between terms without derivatives. During this procedure, it is convenient to introduce two functions:
\be
f_0(\vec x) \equiv i b_0 + a_1^{(1)}V_1(\vec x); \quad f_3(\vec x) \equiv b_3 - a_2^{(1)}V_1(\vec x). \label{8-0}
\ee
In terms of linearly transformed coordinates,
\be
y_2\equiv \alpha x_1 + \beta x_2; \quad y_1\equiv -\alpha x_2 + \beta x_1;\quad \partial_1=\alpha \partial_{y_2}+\beta \partial_{y_1};\quad
\partial_2=\beta \partial_{y_2}-\alpha \partial_{y_1}, \label{8-5}
\ee
with constant complex parameters $\alpha,\, \beta$ defined by coefficients from (\ref{a}), (\ref{b}):
\be
\alpha \equiv a_3^{(1)}+ia_0^{(2)}; \quad \beta \equiv a_3^{(2)}-ia_0^{(1)}, \label{8-55}
\ee
one can check that these functions satisfy the condition:
$$
\partial_{y_2}f_0(\vec x)-\partial_{y_1}f_3(\vec x) = 0.
$$
For this reason, $f_0,\,f_3$ are expressed in terms of one function:
\be
f_0(\vec x) = \partial_{y_1}f(\vec x);\quad f_3(\vec x) = \partial_{y_2}f(\vec x).
\label{xxx}
\ee

After these preparations, simple but rather long calculations of the terms without derivatives in (\ref{NM}) lead to a compact system of nonlinear differential equations:
\ba
(\alpha^2+\beta^2)(\alpha\partial_1+\beta\partial_2)V_-(\vec x) &=& -2V_+(\vec x))(\alpha\partial_1+\beta\partial_2)f(\vec x); \label{10-1}\\
(\alpha^2+\beta^2)(\alpha\partial_2-\beta\partial_1)V_+(\vec x) &=& -2V_-(\vec x)(\alpha\partial_2-\beta\partial_1)f(\vec x); \label{10-2}\\
(\alpha^2+\beta^2)V_+(\vec x) V_-(\vec x) &=& 2(\partial^2_1+\partial^2_2) f(\vec x). \label{10-3}
\ea

Thus, from the intertwining relations (\ref{int}), we have the relations (\ref{abab}) and the system of equations (\ref{10-1}) - (\ref{10-3}) together with (\ref{8-0}). Arbitrary solution for potentials $V_{\pm}$ and function $f$ provides the pair of potentials $V_1,\,V_2$ and intertwining operators $N(\vec x),\, M(\vec x)$ which satisfy (\ref{int}). Nonlinearity of Eq.(\ref{10-3}) and the condition of reality of both potentials for (in general) complex constant parameters made the task nontrivial. To solve this system, one can approach by means of some suitable ansatzes - in particular, by choosing some special values of constant parameters above. For example, it is convenient to deal below with real normalized constants $\alpha,\, \beta,$ i.e. with
\be
a_0^{(1)} = a_0^{(2)}= 0; \quad \alpha^2+\beta^2=(a_3^{(1)})^2 + (a_3^{(2)})^2 = 2. \label{alpha}
\ee

\section{Examples.}

{\bf Example 1.} 

There is a special case when the massless Dirac equation (\ref{dirac}) can be solved analytically without problems: the case with zero external potential $V(\vec x)=0.$ The two-component solution of the corresponding Dirac equation is expressed by arbitrary functions of variables $z=x_1+ix_2 $ or $\bar z=x_1-ix_2:$
\be
\Psi_A = \Psi_A(z);\quad \Psi_B = \Psi_B(\bar z).
\label{Vzero}
\ee
Due to the well known Liouville theorem \cite{complex}, normalizability of such solution is possible only for the restricted regions of the plane with suitably chosen boundary conditions. Physically, this corresponds to the well known Klein paradox \cite{novoselov} - \cite{novoselov-4}, \cite{IN}.

Let us consider just the case:
\be
V_2(\vec x)=0;\quad V_+(\vec x)=V_-(\vec x)= V_1(\vec x)
\nonumber
\ee
with parameters from (\ref{alpha}). It follows from (\ref{10-1}), (\ref{10-2}) that:
\be
V_1(\vec x) = c\cdot\exp{(f(\vec x))},
\label{Vf}
\ee
with real constant $c,$ and therefore, from (\ref{10-3}) - that $f(\vec x)$ must satisfy the nonlinear equation:
\be
(\partial_1^2+\partial_2^2)f(\vec x) = - c^2\cdot \exp{(2 f(\vec x))}.
\label{VVf}
\ee
This equation is familiar in the burning theory, and for real values of $c$ two different solutions are known \cite{polyanin}.

First solution

For the first of them,
\be
f(\vec x)= - \ln{\biggl((\vec x + \vec\tau)^2 + \gamma^2\biggr)} + \ln{(\frac{2\gamma}{c})},
\label{sol-1}
\ee
with arbitrary constants $\tau_1, \tau_2, \gamma$, one obtains:
\be
V_1(\vec x)= 2\gamma\frac{1}{(\vec x + \vec\tau)^2+\gamma^2}.
\label{pot-1}
\ee
Due to generalized intertwining relations (\ref{int}), solutions of the Dirac equation (\ref{dirac}) with this potential can be obtained according to (\ref{solutions}).
It is necessary to use the expressions derived above for coefficients $A_1,\, A_2$ and for function $A(\vec x):$
\ba
&&(A_1\partial_1+A_2\partial_2) = \left(
                \begin{array}{cc}
                  (\alpha +i\beta )\partial + (\alpha -i\beta )\bar\partial & 2\bar\mu\partial  \\
                  2\mu\bar\partial & -(\alpha +i\beta )\partial - (\alpha -i\beta )\bar\partial \\
                \end{array}
              \right);    \label{E-5}\\
&&A(\vec x)=\left(
                \begin{array}{cc}
                  -(if_0+f_3) & \frac{1}{2}V_1(\vec x)(\beta + i\alpha )  \\
                  \frac{1}{2}V_1(\vec x)(\beta - i\alpha ) & -(if_0-f_3) \\
                \end{array}
              \right);  \, \mu\equiv a_1^{(1)} + ia_2^{(1)};\, \bar\mu\equiv a_1^{(1)} - ia_2^{(1)}. \label{E-2}
\ea
Taking into account the simplifying ansatz (\ref{alpha}) and explicit analytical expressions (\ref{sol-1}), (\ref{pot-1}), after the straightforward calculations, one obtains from (\ref{Vzero}) both components of solution with potential $V_1(\vec x)$:

\ba
\Psi_A^{(1)}(\vec x) &=& -(\alpha + i\beta )\frac{\bar z + \bar\tau }{(\vec x + \vec\tau)^2+\gamma^2}\Psi_A(z) + \frac{\gamma (\beta + i\alpha)}{(\vec x + \vec\tau)^2+\gamma^2}\Psi_B(\bar z) + (\alpha + i\beta )\Psi_A^{\prime }(z); \label{E-10}\\
\Psi_B^{(1)}(\vec x) &=& -(\alpha - i\beta )\frac{z + \tau }{(\vec x + \vec\tau)^2+\gamma^2}\Psi_B(\bar z) + \frac{\gamma (\beta - i\alpha)}{(\vec x + \vec\tau)^2+\gamma^2}\Psi_A(z) - (\alpha - i\beta )\Psi_B^{\prime }(\bar z). \label{E-11}
\ea
Here, $z, \bar z$ were defined above, the constants $\tau \equiv \tau_1 + i\tau_2, \, \bar\tau \equiv \tau_1 - i\tau_2$ can be made equal zero by means of translation of $\vec x$, and the constants $\alpha,\, \beta , \mu,\, \bar\mu $ are still arbitrary. The initial solutions for zero potential $\Psi_A(z)$ and $\Psi_B(\bar z)$ can be also chosen as arbitrary functions.

Let us illustrate this example by formulation of the possible boundary problem for this model on an upper half-plane $x_2\geq 0.$ The general boundary conditions \cite{PRL} - \cite{PRL-3} for this domain have the form of linear combination of the components $\Psi_A(x_1, x_2=0)$ and $\Psi_B(x_1, x_2=0):$
\be
a_A\Psi_A^{(1)}(x_1, x_2=0) + a_B\Psi_B^{(1)}(x_1, x_2=0) = 0,
\label{z-1}
\ee
where $a_A, a_B$ are complex constants. It is necessary to find such functions $\Psi_A(z),\, \Psi_B(\bar z)$ in (\ref{Vzero}), that after their substitution into the r.h.s. of (\ref{E-10}), (\ref{E-11}), the condition (\ref{z-1}) will be fulfilled on the line $x_2=0.$ Direct substitution leads to:
\ba
&&(\alpha +i\beta )\bigl[ -(a_Ax_1+ia_B\gamma )\Psi_A(x_1) +a_A(x_1^2+\gamma^2 )\Psi^{\prime}_A(x_1) \bigr] +\nonumber\\
&&+(\alpha -i\beta )\bigl[ -(a_Bx_1-ia_A\gamma )\Psi_B(x_1) -a_B(x_1^2+\gamma^2 )\Psi^{\prime}_B(x_1) \bigr] = 0,
\label{z-2}
\ea
which can be solved by the following choice in (\ref{Vzero}):
\ba
\Psi_A(z)&=&(z^2+\gamma^2)^{1/2}\exp{(i\frac{a_B}{a_A}\arctan(\frac{z^2+\gamma^2}{\gamma }))};
\label{z-3}\\
\Psi_B(\bar z)&=&(\bar z^2+\gamma^2)^{-1/2}\exp{(i\frac{a_A}{a_B}\arctan(\frac{\bar z^2+\gamma^2}{\gamma }))}.
\label{z-4}
\ea

Second solution

The second solution of (\ref{VVf}) is:
\be
f(\vec x)=\frac{1}{2}\ln{\frac{\vec\tau^2}{c^2 \cdot \cosh^2(\vec x\cdot\vec\tau +\gamma^2)}}
\label{sol-2}
\ee
where both components of the constant vector $\vec\tau \equiv (\tau_1,\, \tau_2)$ and $\gamma$ are arbitrary real constants, and
the expression for the partner potential is given by (\ref{Vf}):
\be
V_1(\vec x)= \pm \frac{\mid\vec\tau\mid}{c\cdot \cosh(\vec x\cdot\vec\tau +\gamma^2)}.
\label{pot-2}
\ee
In this case, the solution actually depends on the projection of $\vec x$ along direction $\vec\tau.$ Calculation of the components of solutions of (\ref{dirac}) gives:
\ba
\Psi_A^{(1)}(\vec x)&=&-\frac{(\alpha + i\beta )(\tau_1 - i \tau_2)}{2}\tanh{(\vec x\vec\tau  + \gamma^2)}\Psi_A(z)\pm \nonumber\\
&\pm &\frac{|\vec\tau|(\beta + i\alpha)}{2c\cdot\cosh{(\vec x\vec\tau  + \gamma^2)}}\Psi_B(\bar z) + (\alpha + i\beta )\Psi_A^{\prime }(z);
\nonumber\\
\Psi_B^{(1)}(\vec x)&=&-\frac{(\alpha - i\beta )(\tau_1 + i \tau_2)}{2}\tanh{(\vec x\vec\tau  + \gamma^2)}\Psi_B(\bar z)\pm \nonumber\\
&\pm & \frac{|\vec\tau|(\beta - i\alpha)}{2c\cdot\cosh{(\vec x\vec\tau  + \gamma^2)}}\Psi_A(z) - (\alpha - i\beta )\Psi_B^{\prime }(\bar z).
\nonumber
\ea

{\bf Example 2.} 

Let us choose the case when from the very beginning, the difference $V_-=V_1-V_2$ actually depends only on one direction in the plane:
\be
V_-=V_-(y_1);\quad y_1=\alpha x_1+\beta x_2;\quad \partial_{y_1}=\frac{1}{2}(\alpha\partial_1+\beta\partial_2).
\label{y-1}
\ee
Then, (\ref{10-1}) provides that also $f=f(y_1)$ depends only on $y_1,$ and from (\ref{10-3}), one obtains that not only $V_-,$ but also $V_+$ depend
only on $y_1:$
\be
V_+=V_+(y_1)=-\frac{2f^{\prime\prime}(y_1)}{V_-(y_1)},
\label{y-2}
\ee
as well. Finally, the differential equation (\ref{10-2}) can be integrated providing:
\be
V_+^2(y_1)+2(f^{\prime}(y_1))^2 = 2\nu^2; \quad \nu = Const.
\label{y-3}
\ee
It is convenient now to define new function:
\be
f^{\prime}(y_1)\equiv \nu \cos\varphi(y_1),
\label{y-4}
\ee
so that:
\ba
V_+(y_1) &=& \sqrt{2}\nu\sin\varphi(y_1);\quad V_-(y_1) = \sqrt{2}\varphi^{\prime}(y_1).
\label{y-5}\\
V_{1, 2}(y_1) &=& \frac{1}{\sqrt{2}}(\nu\sin\varphi(y_1) \pm \varphi^{\prime}(y_1)).
\label{yy-2}
\ea
By means of the change of function, the latter equation takes the form of the well known Riccati equation:
\be
\chi^{\prime}(y_1)+\frac{V_2(y_1)}{\sqrt{2}}\chi^2(y_1)-\nu \chi(y_1) + \frac{V_2(y_1)}{\sqrt{2}}=0; \quad \chi(y_1) \equiv \tan\frac{\varphi(y_1)}{2}, \label{yy-4}
\ee
which is solvable \cite{riccati}, \cite{riccati-2} for some kinds of coefficient function $V_2(y_1).$ If one has solution $\chi(y_1)$ for some specific $V_2(y_1),$ the partner potential can be built as:
\be
V_1(y_1)=\frac{2\sqrt{2}\nu\chi(y_1)}{1+\chi^2(y_1)} - V_2(y_1).
\label{yy-5}
\ee
If one knows not only a pair of partner potentials $V_{1, 2}(y_1),$ but also the solutions of Dirac equation for one of these potentials $V_2(y_1),$ the solutions of Dirac equation with the partner potential (\ref{yy-5}) can be built according to prescription (\ref{solutions}).

First solution

The initial massless Dirac equation is easily solved not only for vanishing scalar potential as in Example 1 but also for a constant potential $V(\vec x)=C.$ In this case it is reduced to the homogeneous Helmholtz equation for components of the wave function:
\be
[(\partial_1^2+\partial_2^2)+C^2]\Psi_B(\vec x)=0;\quad \Psi_A(\vec x)=-\frac{i}{C}\partial\Psi_B(\vec x);\quad \partial\equiv\partial_z=\frac{1}{2}(\partial_1-i\partial_2).
\label{y-0}
\ee
Solutions of the homogeneous Helmholtz equation are well known \cite{polyanin-helmholtz}, they depend crucially on the boundary value problem for some domain in the plane and on the chosen system of coordinates.

We use the fact that Eq.(\ref{yy-4}) is solvable for $V_2(y_1)=\sqrt{2}c=const.$ The analytic expressions for solutions $\chi(y_1)$ are different depending on the sign of the constant $(\nu^2-4c^2).$

For the positive $\lambda^2 \equiv (\nu^2-4c^2) >0:$
\be
\chi(y_1)=\frac{1}{2c}\biggl[\nu + \lambda\frac{c_1 e^{\lambda y_1/2}-c_2 e^{-\lambda y_1/2}}{c_1 e^{\lambda y_1/2}+c_2 e^{-\lambda y_1/2}} \biggr],\quad c_{1.2}=const,
\nonumber
\ee
for the opposite sign, $\kappa^2 \equiv (4c^2-\nu^2) >0:$
\be
\chi(y_1)=\frac{1}{2c}\biggl[ \nu - \kappa \tan(\frac{\kappa y_1}{2}+\varphi_0) \biggr];\quad \varphi_0=const;
\nonumber
\ee
and for $\nu^2=4c^2:$
\be
\chi(y_1)=\frac{1}{2c}\frac{\nu y_1 + \nu c_3+2}{y_1+c_3}; \quad c_3=const.
\nonumber
\ee
By means of translations along $y_1,$ these solutions can be simplified correspondingly as:
\ba
\chi(y_1)&=&\frac{1}{2c}(\nu + \lambda\tanh\frac{\lambda y_1}{2}), \quad or \quad \chi(y_1)=\frac{1}{2c}(\nu + \lambda\coth\frac{\lambda y_1}{2});
\nonumber\\
\chi(y_1)&=&\frac{1}{2c}(\nu - \kappa\tan\frac{\kappa y_1}{2});
\nonumber\\
\chi(y_1)&=&\frac{1}{2c}(\nu + \frac{2}{y_1}),
\nonumber
\ea
with the partner potentials defined explicitly by Eq.(\ref{yy-5}).

Second solution

Let us again consider the case when both potentials $V_{1, 2}(\vec x)$ and $f(\vec x)$ depend only on $y_1$ (see Eqs.(\ref{y-1}) -- (\ref{yy-5})) but with another choice for $V_2(y_1)$ which admits solvability of (\ref{yy-4}). Two such solutions have relatively simple form:
\ba
V_2(y_1)&=&\frac{(\nu -1)}{\sqrt{2}\cosh(y_1)};\quad \chi(y_1)=e^{y_1}; \quad \nu < 1; \label{yy-12}\\
V_1(y_1)&=&\frac{2\sqrt{2}\nu e^{y_1}}{\sqrt{1+e^{2y_1}}} + \frac{1-\nu }{\sqrt{2}\cosh(y_1)}; \label{yy-13}
\ea
and
\ba
\widetilde{V}_2(y_1)&=&-\frac{\sqrt{2}}{\cosh(y_1)}; \quad \widetilde{\chi}(y_1)=-\exp{(-y_1)};
\nonumber\\
\widetilde{V}_1(y_1)&=&-\frac{2\sqrt{2}e^{-y_1}}{\sqrt{1+e^{-2y_1}}} + \frac{\sqrt{2}}{\cosh(y_1)}.
\nonumber
\ea
Other solutions of Eq.(\ref{yy-4}) with the same form of $V_2$ exist but they lead to much more complicated expressions for $V_1$.

By means of generalized intertwining relations (\ref{int}) and according to (\ref{solutions}), solutions $\Psi^{(1)}_{A, B}(\vec x)$ of the Dirac equation (\ref{dirac}) with potential (\ref{yy-13}) can be obtained from the solutions $\Psi^{(2)}_{A, B}(\vec x)$ of the Dirac equation with potential (\ref{yy-12}). The matrix differential operator which transforms $\Psi^{(2)}_{A, B}(\vec x)$ to $\Psi^{(1)}_{A, B}(\vec x)$ is:
\ba
&&(A_1\partial_1+A_2\partial_2+A(\vec x)) =
\nonumber\\
&&=\left(
                \begin{array}{cc}
                  2\partial_{y_2}+i(\bar\mu V_2 - f_0) & \bar\mu(\beta +i\alpha)(\partial_{y_1}-i\partial_{y_2})+\frac{1}{2}(\beta + i\alpha)V_+  \\
                  \mu(\beta -i\alpha)(\partial_{y_1}+i\partial_{y_2})+\frac{1}{2}(\beta - i\alpha)V_+ & -2\partial_{y_2}+i(\mu V_2 - f_0) \\
                \end{array}
              \right),\nonumber
\ea
whose constant $\mu$ was defined in (\ref{E-2}).
Fortunately, the components $\Psi^{(2)}_{A, B}$ can be extracted from the papers \cite{peres-5}, \cite{ho-1}, \cite{peres-8}, where solutions of Dirac equation for potential
$V_2(x_1)=\frac{(\nu -1)}{\sqrt{2}\cosh(x_1)}$ of the form analogous to (\ref{yy-12}) (but depending on $x_1$) were obtained explicitly. The Dirac operator $D_1$ in (\ref{dirac}) with potential $V_1(y_1)$
depending only on $y_1$ and the Dirac operator in \cite{ho-1} with potential $V_1(x_1)$ depending only on $x_1$ are connected by an unitary matrix. Therefore, the same unitary operator transforms solutions of \cite{ho-1} into solutions $\Psi^{(2)}_{A, B}(y_1),$ the latter are necessary to obtain $\Psi^{(1)}_{A, B}(y_1)$ according to (\ref{solutions}).

\section{Conclusions.}

It is known that the two-dimensional massless Dirac equation with potentials of different nature plays an important role in the description of graphene and some other materials \cite{novoselov} - \cite{novoselov-4}. Up to now, analytical solutions of such Dirac equation were obtained \cite{peres} - \cite{peres-12} for the restricted class of scalar (electrostatic) potentials by different approaches including methods of SUSY Quantum Mechanics \cite{susy}, \cite{susy-2}. As a rule, these potentials depend only on one coordinate on the plane, and the corresponding one-dimensional part of the wave function is chosen to be normalizable. Meanwhile, the full wave function is non-normalizable due to plane wave multiplier in the second coordinate, this fact is in accordance with the well known Klein paradox \cite{novoselov} - \cite{novoselov-4}. In \cite{IN}, the case of potentials with non-trivial dependence on both coordinates was studied, and solutions (also non-normalizable on the whole plane) were found for a class of such potentials.

In the present paper, the new class of two-dimensional potentials in the massless Dirac equation was considered. These potentials were obtained as solutions of modified intertwining relations with first order matrix intertwining operators. For several specific ansatzes, the solutions of Dirac equation were built analytically. In general, these solutions are also non-normalizable on the whole plane. In particular, for anzatses with dependence on variable along the single line of the plane, the wave functions include the non-normalizable plane wave multiplier along the orthogonal direction like in \cite{susy} - \cite{ho-1}. The wider class of model potentials amenable to solvability of the corresponding Dirac equation will be useful for further study of physically reproducible systems.

In conclusion, one more point must be discussed. Someone may be interested in the question whether there are other solutions of Dirac equation (\ref{dirac}) besides those constructed above from generalized intertwining relations (\ref{int}). In the standard SUSY QM the answer depends on existence of zero modes of the intertwining supercharge operator. If such zero modes exist, they provide the difference between discrete spectra of partner Hamiltonians and give some additional wave functions (see details in \cite{acdi} and \cite{reviews} - \cite{reviews-5}). In the case of present paper, the answer is quite definite. Let us consider the equation Hermitian conjugate to (\ref{int}):
\be
N^{\dagger}D_1 = D_2 M^{\dagger}.
\label{c-1}
\ee
It is clear that generally speaking, it can produce additional solutions of the Dirac equation (\ref{dirac}): an arbitrary zero mode of intertwining operator $M^{\dagger}$ may provide a new solution. This is just analogous to situation in standard SUSY QM. Thus, we must look for solutions $\Psi(\vec x)$ which satisfy simultaneously:
\be
M^{\dagger}\Psi(\vec x)=0;\quad D_1\Psi(\vec x)=0.
\label{c-2}
\ee
We skip for brevity the detailed analysis of the intertwining relations (\ref{int}) together with relations of (\ref{c-2}). The result is the following.
For the existence of such solutions $\Psi(\vec x),$ among other relations, the functions $f_0(\vec x),\, f_3(\vec x)$ given in (\ref{8-0}) and (\ref{xxx}) must satisfy the following equation:
\be
\partial_{y_1}(f_0(\vec x)-if_3(\vec x))=0.
\label{c-3}
\ee
One can check that (\ref{c-3}) is not satisfied in all examples of Section 3. Therefore, no zero modes of $M^{\dagger}$ give new solutions of Dirac equation, and the variety of solutions of (\ref{dirac}) is exhausted by solutions built in Section 3.

\section{Acknowledgments}

The work of M.V.I. was supported by RFBR Grant No. 18-02-00264-a. We are grateful to Dr. D.Candido for information about paper \cite{PRL-3}.


\begin{thebibliography}{}
\bibitem{witten}
E. Witten,
Nucl. Phys. B {\bf 188}, 513 (1981).
\bibitem{darboux}
G. Darboux, Comptes Rendus {\bf 94}, 1456 (1882).
\bibitem{darboux-2}
M.Crum, Quart.J.Math. {\bf 6}, 121 (1955).
\bibitem{schr}
E. Schr\"odinger,
Proc. Roy. Irish Acad. A {\bf 46}, 9 (1940).
\bibitem{schr-2}
L. Infeld, T.E. Hull,
Rev. Mod. Phys.  {\bf 23}, 21 (1951).
\bibitem{reviews}
F. Cooper, A. Khare, U. Sukhatme,
Phys. Rep. {\bf 251}, 267 (1995).
\bibitem{reviews-2}
B. K. Bagchi, {\it Supersymmetry in Quantum and Classical Mechanics} (Chapman, Boca Raton, 2001).
\bibitem{reviews-3}
D.J.Fernandez C,
AIP Conf. Proc. {\bf 1287}, 3 (2010).
\bibitem{reviews-4}
A. A. Andrianov, M. V. Ioffe,
J. Phys. A {\bf 45}, 503001 (2012).
\bibitem{reviews-5}
D. J. Fernandez C, {\it Trends in supersymmetric quantum mechanics}, arXiv:1811.06449 (to be published in "Integrability, Supersymmetry and Coherent States").
\bibitem{nature}
A. A. Andrianov, N. V. Borisov, M. V. Ioffe,
Sov. Phys. JETP Lett. {\bf 39}, 93 (1984).
\bibitem{nature-2}
V. A. Kostelecky, M. M. Nieto,
Phys. Rev. Lett. {\bf 53}, 2285 (1984).
\bibitem{nature-3}
D. K. C. A. Comtet, A. Bandrauk,
Phys. Lett. B {\bf 150}, 159 (1985).
\bibitem{nature-4}
A. A. Andrianov, M. V. Ioffe,
Phys. Lett B {\bf 255}, 543 (1991).
\bibitem{nature-5}
E. Gozzi, M. Reuter, W. D. Thacker,
Phys. Lett. A {\bf 183}, 29 (1993).
\bibitem{nature-6}
A. A. Andrianov, F. Cannata, D. N. Nishnianidze, M. V. Ioffe,
J. Phys. A {\bf 30}, 5037 (1997).
\bibitem{nature-7}
A. A. Andrianov, F. Cannata, J.-P. Dedonder, M. V. Ioffe,
Int. J. Mod. Phys. A {\bf 14}, 2675 (1999).
\bibitem{nature-8}
F. Cannata, M. Ioffe, G. Junker, D. Nishnianidze,
J. Phys. A {\bf 32}, 3583 (1999).
\bibitem{nature-9}
F. Cannata, M. V. Ioffe, A. I. Neelov, D. N. Nishnianidze,
J. Phys. A {\bf 37}, 10339 (2004).
\bibitem{nature-10}
M. V. Ioffe, S. Kuru, J. Negro, L. M. Nieto,
J. Phys. A {\bf 39}, 6987 (2006).
\bibitem{acdi}
A. A. Andrianov, F. Cannata, J.-P. Dedonder, M. V. Ioffe.
Int. J. Mod. Phys. A {\bf 10}, 2683 (1995).
\bibitem{acdi-2}
L. M. Nieto, A. A. Pecheritsin, B. F. Samsonov,
Ann. Phys. {\bf 305}, 151 (2003).
\bibitem{shemyakova}
E. Shemyakova,
SIGMA (Symmetry, Integrability and Geometry: Methods and Applications) {\bf 9}, 002 (2013).
\bibitem{shemyakova-2}
D. Hobby, E. Shemyakova,
SIGMA (Symmetry, Integrability and Geometry: Methods and Applications) {\bf 13}, 010 (2017).
\bibitem{novoselov}
K. S. Novoselov et. al.,
Nature {\bf 438}, 197 (2005).
\bibitem{novoselov-2}
M. I. Katsnelson,
Materials Today {\bf 10}, 20 (2007).
\bibitem{novoselov-3}
A. N. Castro et. al.,
Rev. Mod. Phys. {\bf 81}, 109 (2009).
\bibitem{novoselov-4}
D. S. I. Abergel et. al.,
Advances in Physics {\bf 59}, 261 (2010).
\bibitem{peres}
N. M. R. Peres, E. V. Castro,
J. Phys.: Cond. Matt. {\bf 19}, 406231 (2007).
\bibitem{peres-2}
P. G. Silvestrov, K. B. Efetov,
Phys. Rev. B {\bf 77}, 155436 (2008).
\bibitem{peres-3}
A. Matulis, F. M. Peeters,
Phys. Rev. B {\bf 77}, 115423 (2008).
\bibitem{peres-4}
J. H. Bardarson, M. Titov, P. W. Brouwer,
Phys. Rev. Lett. {\bf 102}, 226803 (2009).
\bibitem{peres-5}
R. R. Hartmann, N. J. Robinson, M. E. Portnoi,
Phys. Rev. B {\bf 81}, 245431 (2010).
\bibitem{peres-6}
C. A. Downing, D. A. Stone, M. E. Portnoi,
Phys. Rev. B {\bf 84}, 155437 (2011).
\bibitem{peres-7}
V. Jakubsky,
Phys. Rev. D {\bf 91}, 045039 (2015).
\bibitem{peres-8}
R. R. Hartmann, M. E. Portnoi,
Phys. Rev. A {\bf 89}, 012101 (2014).
\bibitem{peres-9}
P. Ghosh, P. Roy,
Phys. Lett. A {\bf 380}, 567 (2015).
\bibitem{peres-10}
C. A. Downing, M. E. Portnoi,
Phys. Rev. B {\bf 94}, 165407 (2016).
\bibitem{peres-11}
C. A. Downing, M. E. Portnoi,
J. Phys.: Condensed Matter {\bf 29} 315301 (2017).
\bibitem{peres-12}
C. A. Downing, M. E. Portnoi,
Nature Communications {\bf 8}, 897 (2017).
\bibitem{susy}
B. Midya, D. J. Fernandez C.,
J. Phys. A {\bf 47}, 285302 (2014).
\bibitem{susy-2}
A. Schulze-Halberg, P. Roy,
J. Phys. A {\bf 50}, 365205 (2017).
\bibitem{IN}
M. V. Ioffe, D. N. Nishnianidze,
Mod. Phys. Lett. B {\bf 32}, 1850329 (2018).
\bibitem{complex}
E. T. Whittaker, G. N. Watson,
{\it A course of modern analysis} (4th edition, Cambridge, At the University Press, 1927) Section 5.63.
\bibitem{polyanin}
A. D. Polyanin, V. F. Zaitsev,
{\it Handbook of Nonlinear Partial Differential Equations}
(Second Edition, Chapman and Hall/CRC Press, Boca Raton-London-New York, 2012), 1912 p.p., Section 5.2.1.
\bibitem{PRL}
M. Kharitonov, J.-B. Mayer, E. M. Hankiewicz,
Phys. Rev. Lett. {\bf 119}, 266402 (2017).
\bibitem{PRL-2}
M. T. Ahari, G. Ortiz, B. Seradjeh,
Am. J. Phys. {\bf 84}, 858 (2016).
\bibitem{PRL-3}
D. R. Candido, M. Kharitonov, J. Carlos Egues, E. M. Hankiewicz,
Phys. Rev. B {\bf 98}, 161111 (2018).
\bibitem{riccati}
E. L. Ince,
{\it Ordinary Differential Equations}
(New York: Dover Publications, 1956).
\bibitem{riccati-2}
A. D. Polyanin, V. F. Zaitsev,
{\it Handbook of Exact Solutions for Ordinary Differential Equations}
(second Edition, Chapman and Hall/CRC, Boca Raton, 2003).
\bibitem{polyanin-helmholtz}
A. D. Polyanin, V. E. Nazaikinskii
{\it Handbook of Linear Partial Differential Equations for Engineers and Scientists}
(Second Edition, Updated, Revised and Extended
Publisher: Chapman and Hall/CRC Press, Boca Raton-London-New York, 2016),
1632 p.p., Section 9.3.
\bibitem{ho-1}
C.-L. Ho, P. Roy,
EPL {\bf 108}, 20004 (2014).

\end{thebibliography}
\end{document}